\documentstyle[prl,aps,multicol,epsfig]{revtex}
\begin{document}
\title{Dynamics of Fluctuating Bose-Einstein Condensates}
\author{M.J. Bijlsma and H.T.C. Stoof \\
Institute for Theoretical Physics, University of Utrecht, \\
Princetonplein 5, 3584 CC Utrecht, The Netherlands}
\maketitle
\begin{abstract}
We present a generalized Gross-Pitaevskii equation
that describes also the dissipative 
dynamics of a trapped partially Bose condensed gas.
It takes the form of a complex nonlinear Schr\"odinger
equation with noise. We consider an approximation to this 
Langevin field equation  
that preserves the correct equilibrium for both the
condensed and the noncondensed parts of the gas. 
We then use this formalism to describe the reversible
formation of a one-dimensional Bose condensate,
and compare with recent experiments.
In addition, we determine the frequencies
and the damping of collective modes in this case.
\end{abstract}
\begin{multicols}{2}
The observation of Bose-Einstein condensation 
in ultracold trapped atomic vapors \cite{anderson95,bradley95,davis95}
has offered the possibility to study the
equilibrium and nonequilibrium properties 
of these degenerate gases experimentally and to 
compare the results with {\it ab initio} calculations.
The latter have as an input only the mass and scattering length
of the particular atom of interest,
and the parameters involved in the experimental setup. 
To describe the various equilibrium and nonequilibrium
properties like for example topological excitations, 
relaxation rates, mode frequencies, damping rates and density profiles, 
theories have been developed
at different levels of sophistication. 

At the most elementary level, the
Gross-Pitaevskii equation already captures many 
of the experimentally observed phenomena \cite{dalfovo99}. 
It describes in Hartree approximation the zero-temperature 
dynamics of the condensate,
and has been used to explain and predict many
features of these Bose-condensed systems.
At the next level, the static and dynamic properties
of the noncondensed or thermal part of the gas are to
be included. This can in first instance be done by including
into the Gross-Pitaevskii equation the Hartree-Fock
interaction with the thermal cloud and coupling it 
to an equation for the dynamics of the thermal part of the gas. 
The latter is in good
approximation given by a Boltzmann equation
for the single-particle distribution function
\cite{kirkpatrick85}.   

All these theories describe in essence
only the average dynamics of the gas.
Near the critical region however, fluctuations in the
order parameter are generally much larger
than the average value of the order parameter itself. Therefore,
in this region, it is necessary to include 
fluctuations into a description of the trapped gas.
Also far below the critical temperature,
it can be essential in some cases to include fluctuations
into a description of the dynamics of the gas.
For example, to understand the  
phenomenon of phase diffusion \cite{lewenstein96}, 
one needs to consider fluctuations that
disturb the phase of the condensate.
From a fundamental point of view, 
the desired  dissipative generalization of the 
Gross-Pitaevskii equation should obey 
the so-called fluctuation-dissipation theorem \cite{kadanoff62}. 
This guarantees that both the condensed as well as the noncondensed 
components of the gas will relax to thermal equilibrium.
As a result, the order parameter fluctuates
around its mean value, and the central quantity describing
the dynamics of the condensate is not this mean value, 
but the actual probability distribution of the order parameter. 
With all this in mind, 
a unified theory describing 
the coherent dynamics of the condensate wave function,
the incoherent scattering between
the various components of the gas, as well as the fluctuations
around the average value of the order parameter
has been developed by one of us \cite{stoof97,stoof99}.

The purpose of this Letter is two-fold. First, we show
how the coupled dynamics of the
thermal cloud and the condensate can be
solved in a selfconsistent way, that on the one hand 
leads to the correct
equilibrium distribution of the trapped gas, and at
the same time takes into account
both mean-field effects as well as fluctuations. 
Second, we show that  
fluctuations can be of crucial importance when trying
to understand recent experimental results. 

In general, a theoretical
description of a trapped interacting Bose gas is possible
in terms of the Langevin equation 
\begin{eqnarray}
\label{eq1}
i \hbar {\partial \Phi({\bf x},t) \over \partial t} & = & 
\Bigl[
- {\hbar^{2} \nabla^{2} \over 2 m} + V_{\rm ext}({\bf x}) - \mu
- i R({\bf x},t) 
\\ \nonumber & & 
+ T^{2B} | \Phi({\bf x},t) |^{2} \Bigr]  
\Phi({\bf x},t) +
\eta({\bf x},t) \; .
\end{eqnarray}
Here $\hbar$ is Planck's constant, $m$ is the mass of the
atom, $V_{\rm ext}({\bf x})$ is the external trapping potential,
and $T^{2B}=4 \pi \hbar^{2} a/m$ is the s-wave approximation
to the two-body scattering matrix, with $a$ the scattering length.    
This Langevin equation can be derived
using a field-theoretic formulation of the
Keldysh formalism \cite{stoof99},
and describes the fluctuations as well as the mean-field effects
of both the condensed and the noncondensed parts of the gas.
The derivation requires that the high-energy part of 
the system is sufficiently
close to equilibrium that it can be described as having
a temperature $T$ and a chemical potential $\mu$, and
can play the role of a `heat bath'.  This requirement is 
usually well satisfied and enters 
through the explicit expression
for the imaginary part in our generalized
Gross-Pitaevskii equation, which, if the gas is sufficiently
close to or below the critical temperature, can be approximated by
\begin{eqnarray}
\label{eq2}
i R({\bf x},t) & = & - {\beta \over 4} \hbar \Sigma^{K}({\bf x})
\Bigl[ -{\hbar^{2} \nabla^{2} \over 2 m} + V_{\rm ext}({\bf x}) 
\\ \nonumber & & 
- \mu
+ T^{2B} | \Phi({\bf x},t) |^{2} \Bigr] \; . 
\end{eqnarray}
Here $\beta$ is $1/k_{B} T$, with $k_{B}$ Boltzmann's constant.

If $R({\bf x},t)$ has this particular form, 
the trapped gas will relax 
to equilibrium, because it enforces
the fluctuation-dissipation theorem. 
Note that Eq.~(\ref{eq2}) is always valid for the 
energy levels below the chemical potential, i.e., for the condensate, 
but causes the energy distribution function
for the noncondensed cloud to relax to its `classical' value
$N(\epsilon)=[\beta (\epsilon-\mu)]^{-1}$. 
Therefore, it cannot describe the exponential decay of the 
density of noncondensed atoms at the edges of the thermal
cloud.
Instead, the density of thermal atoms decays algebraically.
In addition, the classical approximation overestimates the average 
number of atoms for the eigenstates above the chemical
potential. Both defects
can be cured  \cite{remark}, but are 
unimportant for the condensate and the 
low-energy part of the thermal cloud,
where most of the atoms reside, and 
for which our theory is intended to be valid.

Finally, the correlations of the Gaussian noise $\eta({\bf x},t)$ 
in Eq.~(\ref{eq1}) are given by 
\begin{eqnarray}
\label{eq4}
\langle \eta^{*}({\bf x},t) \eta({\bf x}',t') \rangle & = & 
{i \hbar^{2} \over 2} \Sigma^{K}({\bf x}) \delta(t-t') 
\delta({\bf x}-{\bf x}') \; ,
\end{eqnarray}
where the average denotes an average over the different
realizations of the noise. 
The general expressions  for $R({\bf x},t)$ and $\Sigma^{K}({\bf x})$
can be found in Ref.\cite{stoof99},
and lead to a Keldysh selfenergy 
$\Sigma^{K}({\bf x})$ equal to
\begin{eqnarray}
\label{eq5}
\hbar \Sigma^{K}({\bf x}) & = & - {4 i [T^{2B}]^2 \over (2 \pi)^{5} \hbar^{6}}
\int d{\bf p}_{1} d{\bf p}_{2} d{\bf p}_{3}
\delta({\bf p}_{1}-{\bf p}_{2}-{\bf p}_{3})
\\ \nonumber & & 
\delta(\epsilon_{1}-\epsilon_{2}-\epsilon_{3})
(1+N_{1})N_{2}N_{3} \; ,
\end{eqnarray}
with $N_{i} \equiv N(\epsilon_{i})$ the Bose distribution
for the eliminated part of the gas,
 and $\epsilon_{i} = {\bf p}_{i}^{2}/2m + V_{\rm ext}({\bf x}) -\mu$.
It determines the strength of the fluctuations 
through Eq.~(\ref{eq4}) and is related
to the damping by means of the fluctuation-dissipation theorem.

At this point, we explain briefly the 
experimental setup we are considering in the 
rest of this Letter. It is inspired
by a recent experiment by Stamper-Kurn 
{\it et.\ al} \cite{stamperkurn98},
and the ideal realization of the
conditions mentioned above.
In the experiment we are considering, a Bose 
gas is trapped and cooled to a temperature above
the transition temperature.
Subsequently, a dimple is created in 
the external trapping potential,
say along the z-axis, by means of optical techniques. 
This dimple is steep enough such that there is only one 
energy level in the potential perpendicular to the
z-axis. We assume the dimple to be well approximated
by a harmonic potential with trapping frequency 
$\omega_{\perp}$. Factorizing the 
wave function $\Phi_{0}({\bf x})=\Phi(x,y)\Phi(z)$, 
the atoms trapped inside this dimple form 
effectively a one-dimensional gas, with an interaction
strength $g = T^{2B}/2 \pi l_{\perp}^{2}$.
Here, $l_{\perp}=\sqrt{\hbar / m \omega_{\perp}}$
is the harmonic oscillator length,
and $\Phi_{0}$ the harmonic groundstate in the dimple.
By changing the depth of the dimple, its lowest energy level 
can become lower than the chemical potential of the noncondensed
three-dimensional gas. If this situation occurs, 
the atoms will condense into this ground state.
Notice that during this process, the noncondensed 
gas in the three dimensional trapping potential 
will remain close to equilibrium, and represents the `heat bath'.

We now turn to a numerical solution of Eq.~(\ref{eq1}) under
these conditions, using a combination of 
well-known techniques \cite{feit}. 
To ensure particle number conservation
in the absence of the term $i R({\bf x},t)$, we
use an implicit 
method that represents the time evolution operator 
$\exp(- i H \delta t/ \hbar)$ as
$(1 + i H \delta t/2 \hbar)^{-1} \times (1 - i H \delta t/2 \hbar)$.
Here, $H= 
- \hbar^{2} \nabla^{2} / 2 m + V_{\rm ext}({\bf x}) - \mu
- i R({\bf x},t) + g |\overline{\Phi({\bf x},t)}|^{2}$,
with  $\overline{\Phi({\bf x},t_{i})}$ the selfconsistent average 
$[\Phi({\bf x},t_{i})+\Phi({\bf x},t_{i}+\delta t)]/2$.  
The numerical method for solving
Eq.~(\ref{eq1}) can now be found from its solution 
$\Phi({\bf x},t_{i}+\delta t)=
\exp(-i H \delta t/\hbar)[
\Phi({\bf x},t_{i}) - (i/\hbar) \exp(-i H t_{i}/\hbar)
\int_{t_{i}}^{t_{i}+\delta t} dt' \exp(i H t'/\hbar) \eta({\bf x},t')]$
by introducing a new noisy variable
$\xi_{i}({\bf x}) = \exp(-i H t_{i}/\hbar) \\
\int_{t_{i}}^{t_{i}+\delta t} dt' \exp(i H t'/\hbar) \eta({\bf x},t')]$.
The correlations of $\xi_{i}({\bf x})$ are
$\langle \xi^{*}_{i}({\bf x}) \xi_{j}({\bf x}') \rangle = 
{i \hbar^{2} \over 2} \Sigma^{K}({\bf x}) \delta_{ij} \delta t 
\delta({\bf x}-{\bf x}') + {\cal O}(\delta t^{2})$. 
The spatial discretization is straightforward.

As a first application, we show in Fig.~1 that in the case of
a noninteracting vapor, the density of the 
harmonically trapped one-dimensional gas relaxes to 
the correct equilibrium given by
$n(z)=\sum_{\alpha} |\phi_{\alpha}(z)|^{2}/\beta(\epsilon_{\alpha}-\mu)$, where
$\epsilon_{\alpha}$ are the energy eigenvalues and $\phi_{\alpha}(z)$ 
the corresponding eigenfunctions of the Hamiltonian. 
The equilibrium is shown for several 
sizes of the spatial mesh, and we see that the 
density distribution converges towards the continuum
limit given by $[2  l_{z}^{2} \beta \hbar \omega_{z}]^{-1/2}
[\beta (V_{\rm ext}-\mu)]^{-1/2}$ 
where $l_{z} = \sqrt{\hbar / m \omega_{z}}$ is
the harmonic oscillator length along the $z$-direction,
only for rather small mesh sizes.
This is caused by the relatively large contribution of high-energy states
due to the classical behavior of the thermal cloud. 
We emphasize, however, that
given a certain mesh size, the correct equilibrium corresponding
to that particular discretization is reproduced numerically.
Moreover, it is important to realize that if we do not include the noise,
the density of the gas would be zero. 
Hence, it is clear that the fluctuation-dissipation 
theorem ensures that 
the noise and the imaginary term
in Eq.~(\ref{eq1}) 
cooperate in order to occupy the energy
levels thermally. Thus, including fluctuations is crucial
in treating the effects of the thermal cloud.

\begin{figure}[h]
\epsfig{figure=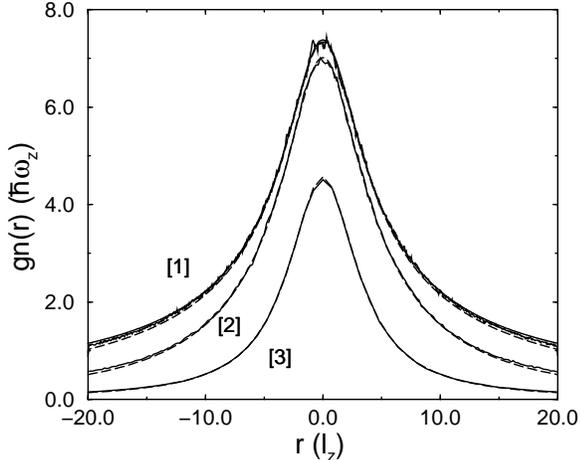}
\caption{
\label{fig1}
\narrowtext
Stationary density profile
of a noninteracting gas 
above the critical temperature. 
The time step
used is $\Delta t=0.002 \; \omega_{z}^{-1}$, 
the discretization $\Delta x$
used for the different lines is
[1]: $\Delta x=0.05 \; l_{z}$,
[2]: $\Delta x=0.2 \; l_{z}$, and
[3]: $\Delta x=0.8 \; l_{z}$.
Also shown are the 
predicted
classical density profiles 
for curves [1], [2], and [3] (dashed lines),
and the continuum result (solid line).
The chemical potential of the three
dimensionally trapped gas is $\mu=-30 \; \hbar \omega_{z}$, 
and its temperature $T=400$ nK. The trapping frequencies 
are $\omega_{\rm z}=2\pi \times 13$Hz and 
$\omega_{\rm r}=2\pi \times 20$Hz \protect\cite{stamperkurn98}. 
The trapping frequency of the
dimple is $\omega_{\perp}=2 \pi \times 500$Hz, and
the ground state eigenvalue $\epsilon_{0}=-25 \; \hbar \omega_{z}$.
The atom considered is sodium, with a scattering
length $a \approx 2.75$ nm. 
}
\end{figure}

Second, we calculate the damping of the breathing mode
in a trapped one-dimensional Bose gas below the 
critical temperature. Shown in Fig.~2 is $g$ times
the value of the density $n({\bf 0})$ in the
center of the one-dimensional trap.
First, the gas relaxes towards equilibrium 
at an effective chemical potential $\mu-\epsilon_{0}=30 \; \hbar \omega_{z}$,
where $\epsilon_{0}$ is the eigenvalue 
of $\Phi_{0}(x,y)$ in the dimple. 
The fact that the value of $g n({\bf 0})$ does not
relax to $30 \; \hbar \omega_{z}$ can be explained by 
realizing that mean-field effects are included
in our formalism, because $\Phi(z,t)$ describes
both the condensed and the noncondensed
parts of the one-dimensional gas. This implies that the chemical potential
for the condensed part of the gas is effectively
lowered by the interaction with the noncondensed
part of the gas.
After allowing the gas to equilibrate, 
at $t=0.61$ s the trapping frequency
is instantaneously changed to $\omega'_{z}=0.95 \; \omega_{z}$,
and the gas starts to oscillate.
The frequency $\omega$ and the damping rate $\gamma$
of the oscillation are found by fitting
to $n({\bf 0},t)=n_{\rm eq}({\bf 0})+\delta n({\bf 0}) 
\exp(-\gamma t) [1-\cos(\omega t)]$,
which describes both the relaxation of the density to 
its new equilibrium, as well as the excitation and damping 
of the oscillation. The results are 
$\omega = 136.84$ s$^{-1}$, which is close to the
result of $\sqrt{5/2} \omega_{z} = 129.1$ s$^{-1}$
expected from the Gross-Pitaevskii equation \cite{stringari}, 
$\gamma = 12.3$ s$^{-1}$, 
$g \delta n({\bf 0}) = 0.84 \; \hbar \omega_{z}$, and 
$g n_{\rm eq}({\bf 0})=28 \; \hbar \omega_{z}$.
Note that the damping is in principle caused both by 
the collisions with the reservoir of thermal atoms in the
three-dimensional trapping potential,
which is described by $\Sigma^{K}({\bf x})$, 
as well as by the nonlinear term in the
Langevin equation which induces damping upon averaging over
the different realizations of the noise.
The latter includes collisional, as well as Landau damping. 

In Fig.~3, a snapshot of
the density profile is shown at
several times during the initial growth of
the condensate. These are taken from the 
simulation presented in Fig.~2. As expected, the condensate
shows up as a peak in the density profile.
Also shown are the Thomas-Fermi solution
$[\mu - V_{\rm ext}(z)]/g$ for the condensate density
and the noninteracting result for the noncondensed
part of the gas which diverges at the critical
point where the effective chemical potential
becomes equal to zero. This is due to the fact that
in one dimension, mean-field theory completely fails near the
critical temperature. Fig.~3 shows that far
enough from the condensate, the noninteracting
result is obtained. Again, in the center of the trap
one obtains the Thomas-Fermi result
only approximately due to interactions.
\begin{figure}[h]
\epsfig{figure=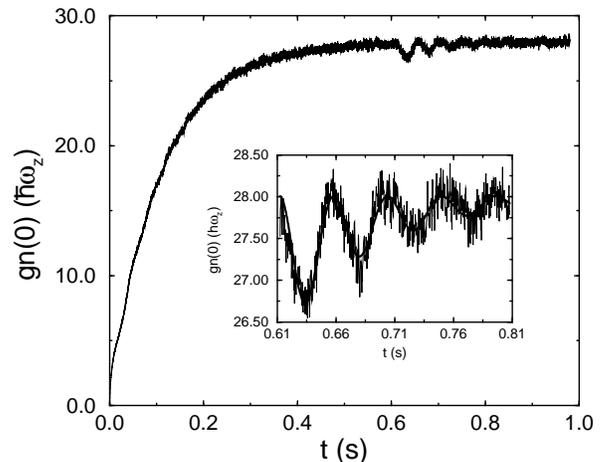}
\caption{
\narrowtext 
\label{fig2}
Oscillations~of~the~trapped~partially Bose-condensed gas, and
their damping, after a small and instantaneous change 
of the trapping frequency $\omega_{z}$ at $t=0.61$ s. 
Here, we used $\Delta t=0.002 \; \omega_{z}^{-1}$, 
$\Delta x=0.05 \; l_{z}$, $\epsilon_{0} = -60 \hbar \omega_{z}$,
and $\omega_{\perp}=2 \pi \times 1200$Hz. 
The remaining parameters are the same as in Fig.~1.
}
\end{figure}

Finally, we consider the reversible formation of
a one-dimensional Bose condensate, similar
in spirit to the experiments by Stamper-Kurn {\it et. al}
\cite{stamperkurn98}. 
The dimple
perturbing the external trapping potential 
is now oscillating in such a way, that the lowest
energy level in the dimple crosses the chemical
potential of the three dimensionally trapped gas several times
according to $\epsilon_{0} = \mu + \mu \sin(\omega t)$,
with $\omega=2 \pi$ s$^{-1}$ and $\mu = -30 \hbar \omega_{z}$.
The calculations have been done with and without
noise. The results show that without noise, the 
condensate evaporation and growth cannot be described properly.
Indeed, in that case, our findings depend strongly on the initial
conditions. Moreover, the periodic growth occurs with decreasing amplitude.  
\begin{figure}[h]
\epsfig{figure=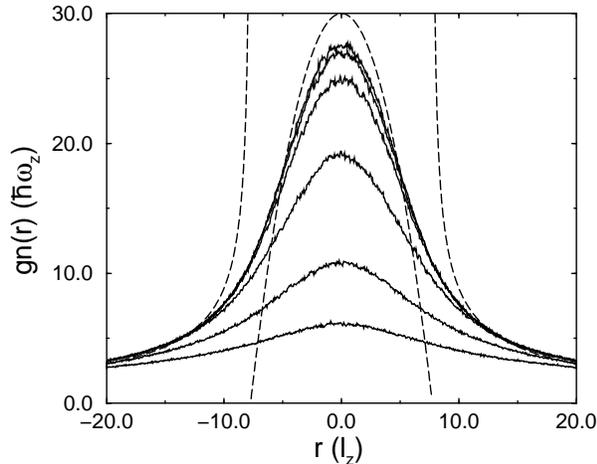}
\caption{
\label{fig3}
\narrowtext 
This figure shows the growth of
the density profile 
of the trapped partially Bose-condensed gas
for the simulation of Fig.~2. 
The snapshots are taken at
$t=2 \; \omega_{z}^{-1}$, 
$t=4 \; \omega_{z}^{-1}$,
$t=10 \; \omega_{z}^{-1}$,
$t=20 \; \omega_{z}^{-1}$,  
$t=30 \; \omega_{z}^{-1}$, and
$t=40 \; \omega_{z}^{-1}$.
The dashed lines show the equilibrium mean-field results 
in and outside the condensate.
}
\end{figure}
To describe the growth cycles correctly, one needs
to include fluctuations into the generalized Gross-Pitaevskii
equation, which is to be expected since
we are at times in the critical region. 
Notice that we
quantitatively reproduce the lagging behind of the condensate
as observed in an experiment performed with an
essentially three-dimensional dimple. We observed 
a lagging behind of roughly $0.1$s, whereas in experiment
$0.07$s was measured \cite{stamperkurn98}.

\begin{figure}[h]
\epsfig{figure=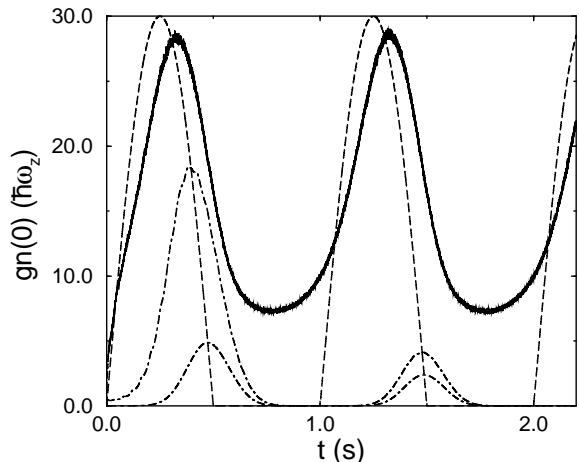}
\caption{
\label{fig4}
\narrowtext 
Reversible formation
of a condensate (solid line) because of a sinusoidal 
time dependence of the chemical potential 
in the dimple (dashed line).
The time step
used was $\Delta t=0.003 \; \omega_{z}^{-1}$, 
the discretization $\Delta x=0.05 \; l_{z}$,
and $\omega_{\perp}=2\pi \times 1200$Hz.
The other parameters used are the same as in Fig.~1.
The two dot-dashed lines correspond to simulations
without noise, with $10$ (top) and $0.1$ (bottom) 
particles in the initial harmonic oscillator ground state.
}
\end{figure}
In conclusion, we have shown that our generalized
stochastic Gross-Pitaevskii equation 
consistently takes into account the 
dissipative dynamics of a trapped partially 
Bose-condensed gas, and describes 
the equilibrium density profile, condensate growth, 
coherent dynamics, and damping in a unified way. 
In our opinion, the one-dimensional experiment 
considered here would be ideal for a detailed
comparison between theory and experiment in
the problem of condensate growth,
because the three-dimensional 
cloud remains in equilibrium.   
In addition, a numerical study of 
the two-dimensional case would be of interest 
and might be used to investigate
for example the dissipative dynamics and 
the formation of vortices in a rotating Bose gas.

\end{multicols}

\end{document}